# Enhanced brightness and photostability of cyanine dyes by supramolecular containment

Hari S. Muddana,[1,3]* Samudra Sengupta,[2] Ayusman Sen,[2] and Peter J. Butler[1]*

[1]Department of Bioengineering, The Pennsylvania State University, University Park, PA 16802

[2]Department of Chemistry, The Pennsylvania State University, University Park, PA 16802

[3]Current address: Skaggs School of Pharmacy and Pharmaceutical Sciences, University of California, La Jolla, CA, 92093

*Corresponding author: hmuddana@ucsd.edu, pbutler@psu.edu

ABSTRACT

Ultrasensitive detection and real-time monitoring of biological processes can benefit significantly from the improved brightness and photostability of the popular organic dyes such as cyanines. Here, using a model cyanine dye, Cy3, we demonstrate that brightness and photostability of the dye is significantly altered when trapped in a molecular container, e.g. cucurbit[n]urils (CB[n]) and cyclodextrins (CD). Through computational modeling, we predicted that Cy3 forms a stable inclusion complex with three different hosts, CB[7], β-CD, and methyl-β-CD, which was further confirmed by single-molecule diffusion measurements using fluorescence correlation spectroscopy. The effect of supramolecular encapsulation on Cy3's photophysical properties was found to be highly host-specific. Up to three-fold increase in brightness of Cy3 was observed when the dye was trapped in methyl-β-CD, due to an increase in both the dye's absorption and quantum yield. Steady-state and time-resolved spectroscopy of the various complexes revealed that host's polarizability and restricted mobility of the dye in the host both contribute to the observed increase in molecular brightness. Furthermore, entrapment of the dye molecule in CDs resulted in a marked increase in the dye's photostability, whereas the dye degraded more rapidly in CB[7]. These results suggest that the changes in photophysical properties of the dye afforded by supramolecular encapsulation are highly dependent on the host molecule. The reported improvement in brightness and photostability together with the excellent biocompatibility of cyclodextrins makes supramolecular encapsulation a viable strategy for routine dye enhancement.

Recent advances in super-resolution imaging have extended the spatial resolution limit of fluorescence microscopy down to tens of nanometers.[1,2] Enabling these techniques for effective use in biomolecular imaging at single-molecule resolution requires substantial improvement in sensitivity and stability of the fluorescent probes.[3] Organic dyes such as cyanines, rhodamines, and fluorescein, are among the most widely used fluorescent probes for biomolecular imaging, owing to their size, specificity, and solubility. However, these fluorescence markers often exhibit rapid photobleaching, low quantum yields, or blinking.[4-6] Dye-doped nanoparticles alleviate some of these problems by shielding the dye from interacting with the solvent or other dissolved reactive species such as singlet oxygen.[7-13] Nevertheless, these nanomaterials are often faced with a different set of challenges including cytotoxicity,[14-16] immune response,[17-19] biodistribution/clearance,[20] aggregation,[21,22] and leakage of cargo[11]. Also, total encapsulation of the dye results in a loss of specificity, and therefore requires further functionalization for targeted labeling.

An alternative to nanoencapsulation is the supramolecular host-guest complexation using molecular containers such as cyclodextrins (CDs) and cucurbiturils (CBs).[23-25] These macrocyclic molecular containers are known to bind a wide variety of guests including ions, fluorophores, and drugs, with high affinities,[26,27] and exhibit low cytotoxicity[28-30]. Often, supramolecular encapsulation alters the dye's spectral properties due to a change in the local microenvironment, such as an increase or decrease in local polarity and viscosity.[23,24,31] This phenomenon is frequently used in detection of other competing guests through indicator displacement assay.[32,33] The host-induced spectral changes can also be used to fine-tune the absorption and emission wavelengths of the dye. However, it has to be noted that supramolecular complexation can sometimes have a detrimental effect on the dye's brightness and photostability, depending on the chemical nature of the host and dye molecules.[23] Another advantage with complexation is that it helps to disperse poorly soluble dyes and prevents aggregation.

In this study, we used experimental and computational methods to assess the mechanisms by which supramolecular encapsulation of sulfoindocyanine dye by CDs and CBs alter the dye's photophysical properties. Cyanine was chosen as the dye-of-interest because of its widespread use in labeling proteins and DNA, photodynamic therapy, microarrays, and optical recording.[34,35] Also, indocyanine green, a member of the cyanine family that excites at near-infrared wavelengths, is currently an FDA-approved dye used in many clinical diagnostic applications.[36] Specifically, we studied the widely used Cy3-maleimide dye, hereafter simply referred to as Cy3 (Figure 1), since its peak absorption wavelength lies within the visible spectrum at 550 nm. Several CDs and CBs were first computationally screened as viable hosts for trapping Cy3. Hosts that showed promise for Cy3 binding were further experimentally tested for complex formation using fluorescence correlation spectroscopy (FCS). Detailed insights into the mechanism of the dye's photophysics were obtained through steady-state (UV-Vis) and time-correlated single-photon counting (TCSPC) measurements. Finally, we determined the effect of supramolecular encapsulation on the photostability of Cy3, and elucidated the underlying mechanism.

**RESULTS AND DISCUSSION**

**Host selection though computational modeling.** Seven hosts, including three cyclodextrins (α-CD, β-CD, and methyl-β-CD) and four cucurbiturils (CB[5], CB[6], CB[7], and CB[8]), were initially selected as potential hosts for Cy3. Through flexible molecular docking of Cy3 to each of these hosts,[37] we found that the cavity sizes of α-CD, CB[5], and CB[6] are too small to form an inclusion complex with Cy3, and therefore ruled out these hosts from further analysis. For the rest of the hosts, we further optimized the geometry of the docked conformation using a more accurate semi-empirical quantum-mechanical energy model, PM6-DH2.[38,39] Lowest energy (i.e. most stable) conformations of the host-Cy3 complexes thus identified are shown in Figure 1. The headgroup chromophore region of Cy3 was found to be docked inside the host's cavity, in all the cases. Based on these lowest-energy conformations, we infer that Cy3 binding to CBs is primarily mediated through hydrophobic interactions, whereas binding to CDs is mediated by both hydrophobic interactions and hydrogen bonding between Cy3's sulfonate groups and the hydroxyl groups at CD's portal rims. Not only do CB-Cy3 complexes lack the hydrogen bonding observed in CDs, the sulfonate groups of Cy3 are positioned away from the carbonyl portals of CB, which is likely due to the strong negative electrostatic field created by the carbonyl groups at the portal rims.

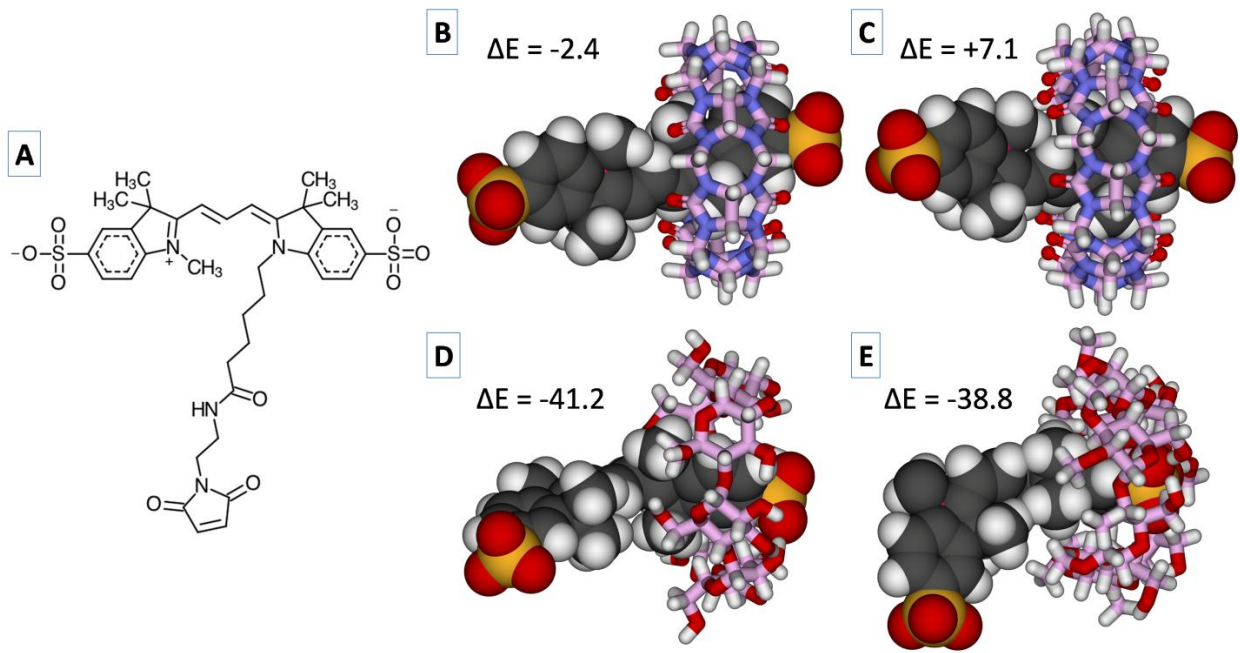

**Figure 1.** (A) Chemical structure of Cy3-maleimide. (B&C) Structures of Cy3 complexed to CB[7] and CB[8], respectively. (D&E) Structures of Cy3 complexed to β-CD and methyl-β-CD, respectively. For clarity, only the headgroup chromophore region of the dye is shown in all the figures. Interaction energy, $\Delta E = E(complex) - E(host) - E(Cy3)$, is also shown for each figure. Color legend: white – hydrogen, grey – carbon (dye), pink – carbon (host), blue – nitrogen, red – oxygen, and yellow – sulfur.

Binding affinities of the different host-Cy3 complexes were estimated by computing the interaction energies using a semi-empirical quantum-mechanical energy model (shown in Figure 1).[40,41] While solvent effects were accounted for by using a continuum solvation model, entropic effects were neglected. Binding of Cy3 to β-CD and methyl-β-CD was highly favored with interaction energies of approximately -40 kcal/mol. On the contrary, CB[7] and CB[8] showed much weaker interaction with Cy3 (Figure 1), with interaction energies of -2.4 and +7.1 kcal/mol, respectively. This result suggests that CDs have a higher affinity to Cy3 than CBs, and CB[7] have a slightly better affinity than CB[8]. CBs are well known to exhibit strong binding affinity towards cationic and neutral guests.[26] Given that Cy3 carries a net negative charge, the weaker interaction predicted for Cy3 and CBs was expected. As host-guest binding is generally entropically unfavorable,[40,42] the positive interaction energy (unfavorable), +7.1 kcal/mol, between CB[8] and Cy3 indicates that Cy3 will not form an inclusion complex with CB[8]. The unfavorable interaction between the sulfonate groups of Cy3 and the carbonyl portals of CB might have been overcome by the strong hydrophobic interactions in CB[7], but not in CB[8]. Together, these results suggested that CDs are likely to have a stronger influence on the photophysical behavior of Cy3 compared to CBs, as they bind more tightly to the chromophore region of Cy3.

**Single-molecule detection of host-Cy3 complexes.** Hosts that were computationally identified to interact favorably with Cy3, that is β-CD, methyl-β-CD and CB[7], were chosen for further experimental characterization. To confirm that these hosts form supramolecular complexes with Cy3, we measured the diffusion coefficient and thereof hydrodynamic radius of Cy3 in presence of the different hosts in water using the FCS method. Autocorrelation traces for free-Cy3 and the various complexes are shown in Figure 2A. A diffusion coefficient of 241 ± 12 μm$^2$/s was recorded for free-Cy3 in water, which corresponds to a hydrodynamics radius of 8.1 ± 0.5 Å. The diffusion coefficient of Cy3 in the presence of CB[7], β-CD, and methyl-β-CD was measured to be 206 ± 19, 180 ± 13, and 155 ± 4 μm$^2$/s, respectively. The corresponding hydrodynamic radii were 9.5 ± 0.9, 10.9 ± 0.8, and 12.6 ± 0.3 Å, respectively (shown in figure 2B). A significant decrease (increase) in diffusion coefficient (hydrodynamic radius) of Cy3 in the presence of different hosts confirms that the dye is being trapped by the host molecules.

As the host molecule forms a single-atom thick layer around the dye (as shown in Figure 1), we expected the hydrodynamic radius of host-dye complexes to differ from that of the free-dye by only few Angstroms and to be independent of the host. However, the hydrodynamic radius of the dye varied significantly between the various hosts. Despite our attempt to saturate host-dye binding by using a low concentration of dye and high concentration of the host, the binding reaction may not have been saturated, in which case, the measured diffusion coefficient would be an ensemble average of both the bound and unbound populations of the dye. A larger radius would then correspond to a higher binding affinity, and vice versa. Cy3 complexes with CDs do exhibit larger hydrodynamic radii compared to those with CBs, in accord with their predicted binding affinities. Alternatively, the observed changes in hydrodynamic radius could also arise from the differences in hydration structure.

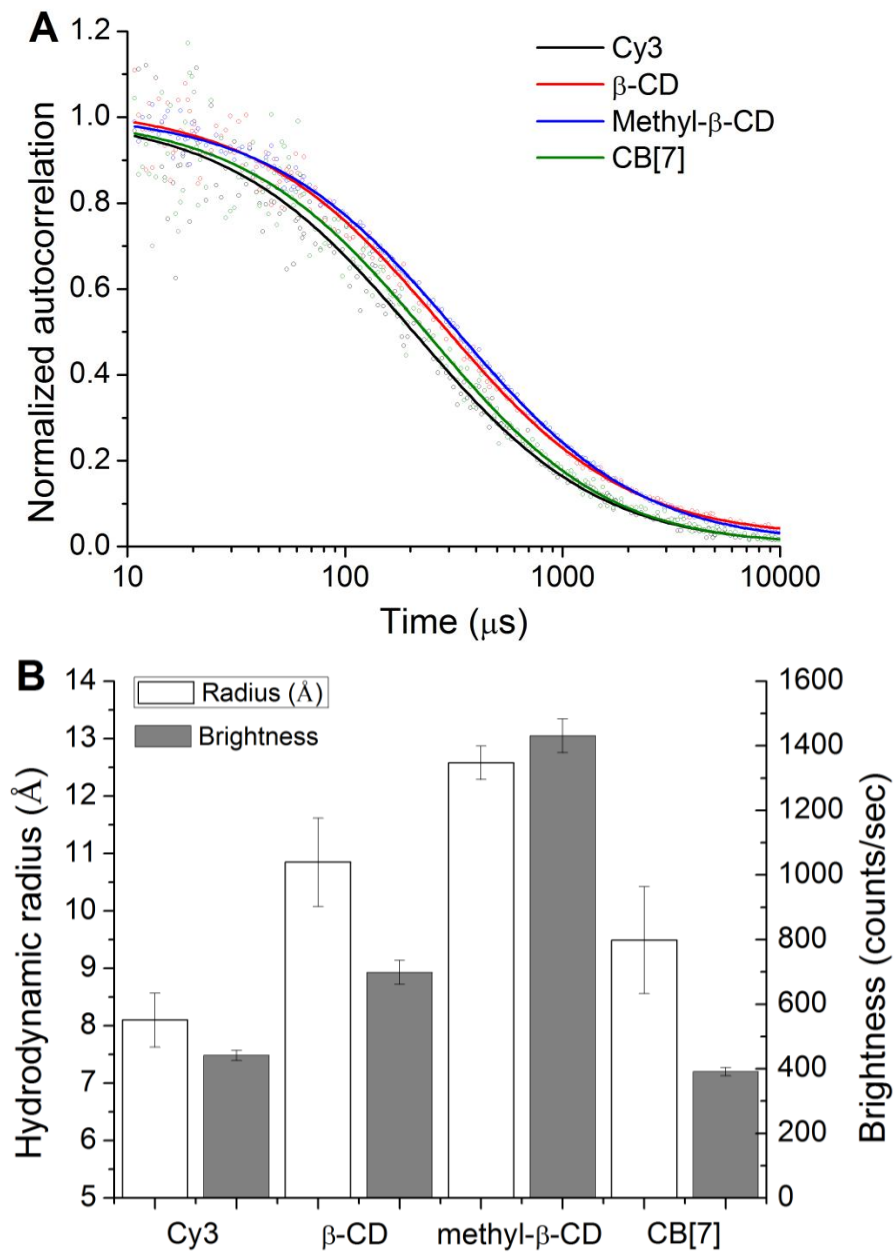

**Figure 2.** (A) Normalized fluorescence autocorrelation curves of free Cy3 and various host-Cy3 complexes determined using time-correlated single photon counting. Solid lines represent theoretical fits to 3D diffusion model. (B) Hydrodynamic radius and molecular brightness (photons emitted per second per molecule) of Cy3 in free and complexed forms. Error bars represent standard deviations (n = 5).

**Molecular brightness enhancement.** Supramolecular complexes of Cy3 exhibited significantly higher brightness compared to the free dye (Figure 2B). Molecular brightness was quantified as the number of photons emitted per second per molecule.[13] Mean fluorescence intensity of the sample was determined from the raw fluorescence intensity traces obtained from the TCSPC instrument by dividing the total number of photons by the total duration of collection, and the average number of molecules in the observation volume was obtained from FCS.[43] Brightness of Cy3 was enhanced by a factor of 1.6 and 3.2 when bound to β-CD and methyl-β-CD, respectively. On the other hand, a slight decrease (10%) in the dye's brightness was observed when bound to CB[7]. While unexpected, it was interesting to find that methylation of β-CD had an additional two-fold enhancement in the brightness of Cy3, compared to β-CD, which is likely due to higher binding affinity of Cy3 to methyl-β-CD or to an additional increase in quantum yield and/or absorption. In any case, this enhancement in brightness is comparable to the 4.5 fold enhancement reported for a single Cy3 molecule when completely encapsulated in a calcium phosphate nanoparticle.[13]

To delineate the underlying mechanism for the observed enhancement in brightness, or lack thereof, we measured the absorption/emission spectra of free-dye and the various complexes (Figure 3). The absorption and emission spectra of Cy3 changed notably when complexed to CDs and CBs, confirming that the dye is in fact bound to these hosts with its headgroup chromophore region in the host cavity, as predicted by the computational model (Figure 1). A consistent red shift in the absorption and emission peaks was observed upon encapsulation, irrespective of the host. Similar bathochromic shifts were reported for several dyes upon forming supramolecular complexes with CDs and CBs,[24] and is generally characteristic of the low polarity of host's cavity region. The oscillator strength (integrated absorption spectrum) of Cy3 increased in the case of CDs, and decreased in CB[7]. These changes in oscillator strength may be explained by the differences in polarizability of the various hosts. Previously, Nau *et al.* have experimentally determined the polarizability of various molecular containers,[25,44] and have shown that polarizability of CB[7] is significantly lower compared to that of water, which explains the decrease in oscillator strength of Cy3 when complexed to this host. On the contrary, polarizability of β-CD is similar to that of water, and that of methyl-β-CD is significantly higher than that of water, which is also consistent with the observed increase in oscillator strength of Cy3 in these hosts.

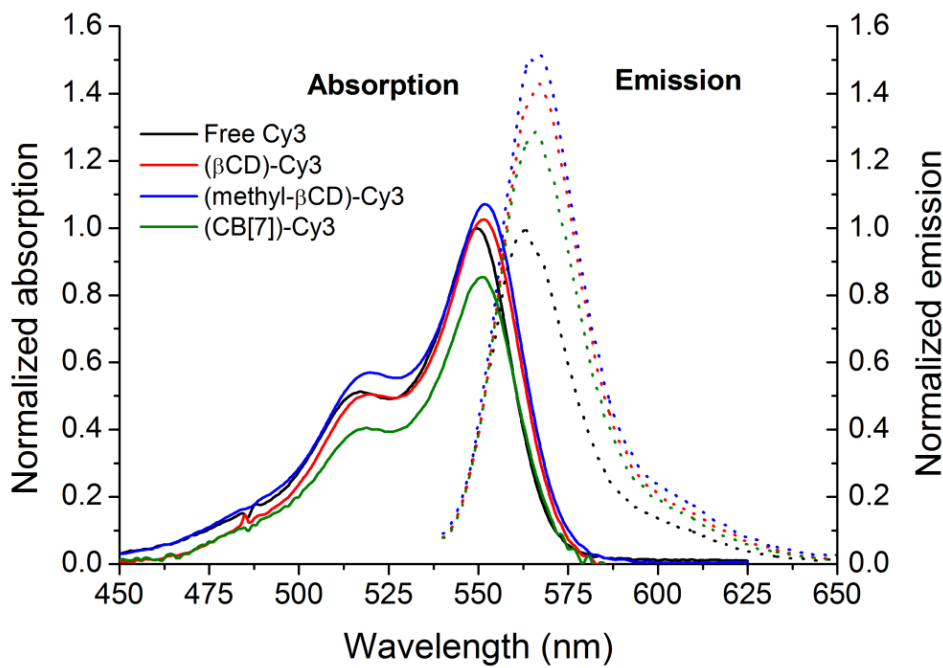

**Figure 3.** Normalized absorption and emission spectra of Cy3 and various host-Cy3 complexes in water.

A concomitant increase in quantum yield of Cy3 is also observed in all the complexes. Quantum yield of Cy3 increased by 10%, 8%, and 14%, when complexed to β-CD, methyl-β-CD, and CB[7], respectively. An increase in both the absorption intensity and the quantum yield of Cy3 in CDs, explains the observed large increase in molecular brightness. In contrast, the increase in quantum yield of Cy3 in CB[7] is compensated by the associated decrease in oscillator strength, resulting in no significant change in molecular brightness. The increase in quantum yield of Cy3 can be attributed to a decrease in the non-radiative decay rate.[45,46] Previously, we and others have shown that trans-cis photoisomerization of the central methine bridge is the primary non-radiative decay pathway of Cy3 from the excited state.[13,45-47] Confinement of the dye molecule through encapsulation results in a significant decrease in the non-radiative decay rate, enhancing the brightness of the dye.[13,47] To test this hypothesis, we measured the fluorescence lifetime of the Cy3 in the free and complexed forms (Figure 4A). Non-radiative decay rates of the dye were calculated from quantum yield and lifetime (Figure 4B). The non-radiative decay rate of the dye was significantly decreased when bound to either of the cyclodextrins, indicating that encapsulation by CDs significantly hinders the mobility of the dye, which is reasonable considering that these hosts exhibit tighter binding affinity to the dye. On the other hand, the non-radiative decay rate of Cy3 was only slightly decreased when bound to CB[7], which can also be attributed to the weaker binding affinity of this host. In summary, steady-state and time-resolved spectroscopy of the dye in free and complex states provided detailed mechanistic insights into the underlying changes in photophysics of Cy3 upon complexation.

**Photostability.** Photodegradation of the dye, i.e. irreversible degradation of the dye from emitting fluorescence, is another important property to be considered in choosing dyes for practical applications. To assess the effect of complexation on photostability of Cy3, we monitored the fluorescence intensity of the free dye, and the different complexes over the course of continuous illumination under epi-fluorescence (Figure 5). Complexation of the dye to β-CD and methyl-β-CD induced a significant increase in photostability of the dye, while CB[7] induced a significant decrease in dye's stability. However, in contrast to the photostability afforded by nanoparticle encapsulation,[12,13] the increase in photostability due to CD complexation is considerably less. In general, the key mechanism through which the dye undergoes bleaching is photooxidation, wherein the dye in triplet-excited state reacts with the ground-state oxygen to form oxygen radicals, which in turn reacts with the dye causing permanent damage to the dye.[48] The rate of photooxidation is dependent on both the rate of triplet-state formation and the rate of reaction with oxygen. Increase in photostability due to complexation or encapsulation is generally attributed to the shielding of the dye from reactive species in solution.[49,50] Therefore, particulate encapsulation generally results in increased photostability of dye molecules.[7,8,13] Given that we observe an increase in photostability in CDs and a decrease in CB[7], we suspect that the underlying mechanism is different from the "solvent shielding" effect. Also, as seen in Figure 1, complexation of Cy3 with CDs and CBs affords only partial protection to the dye, with the dye being largely solvent-accessible. On the other hand, it has been previously shown that CB[7] encapsulation leads to increased triplet state formation, and CDs lead to decreased triplet state formation.[50,51] Based on these studies, we propose that the triplet state formation and not solvent accessibility is the mechanism underlying the changes in photostability due to supramolecular encapsulation. The precise reason for these host-specific changes in triplet state formation is yet to be investigated.

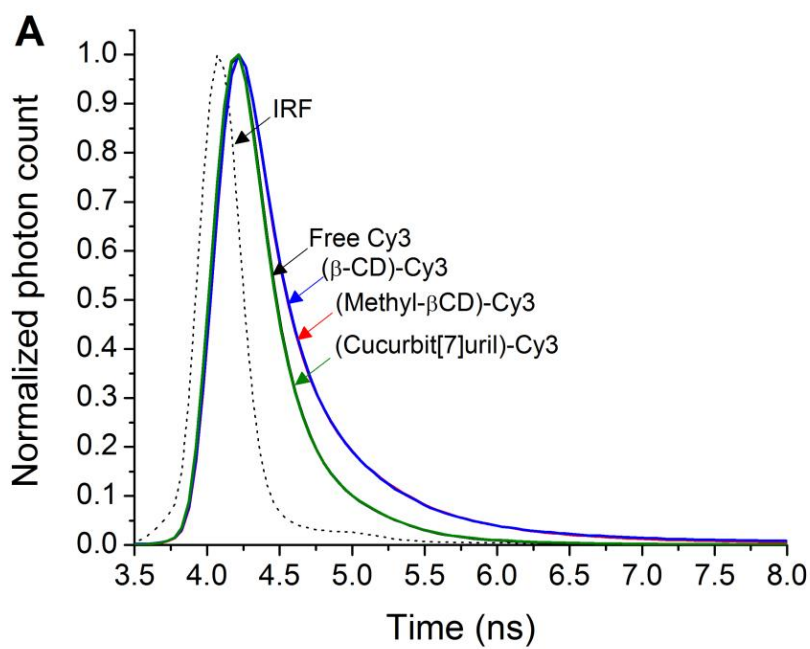

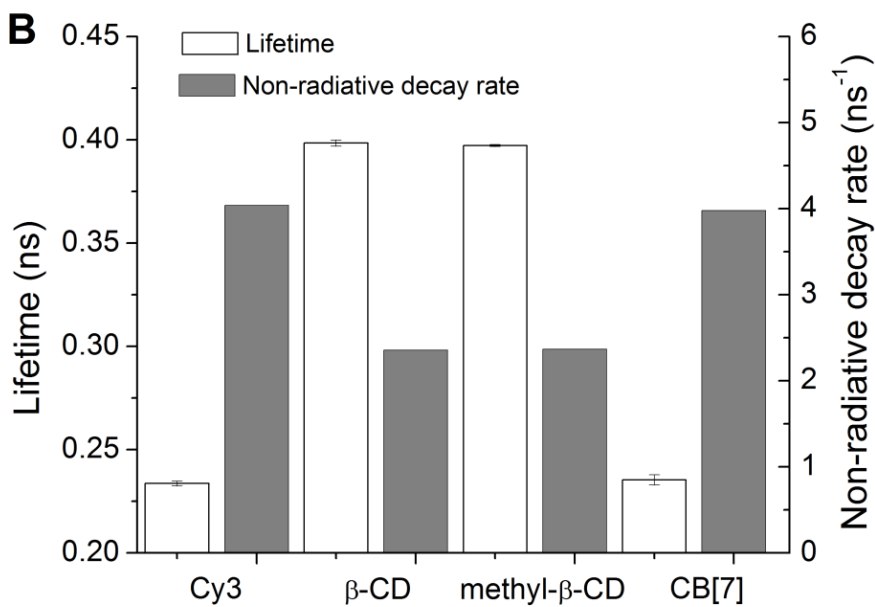

**Figure 4.** (A) Normalized fluorescence lifetime decay curves of Cy3 in free and complexed forms. The dotted line represents instrument response function (IRF). (B) Average fluorescence lifetime and non-radiative decay rates of Cy3 in free and complexed forms (n=5).

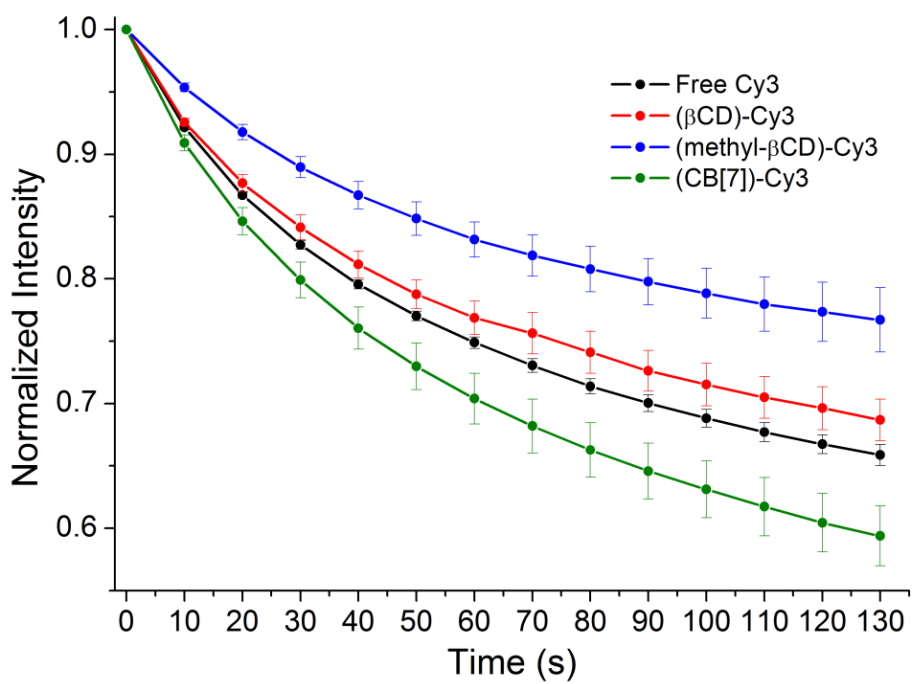

**Figure 5.** Normalized fluorescence intensity of free Cy3 and various host-Cy3 complexes as a function of time, observed during continuous illumination under epifluorescence microscope (n=3).

In summary, we demonstrate that brightness and photostability of a cyanine dye, Cy3, is enhanced significantly by supramolecular encapsulation. Complex formation was confirmed at the single-molecule level using FCS. The underlying mechanisms for the enhanced brightness of complexed dye were elucidated using steady-state and time-resolved fluorescence spectroscopy. Encapsulation-mediated increase in brightness of CD-Cy3 complexes is attributed to the higher binding affinity, better polarizability of the host, and restricted mobility of the dye within the host cavity. Furthermore, supramolecular encapsulation lead to significant changes in photostability of the dye, and we elucidate that triplet state formation is the underlying reason. The enhancement of Cy3 due to supramolecular complexation along with the excellent biocompatibility of cyclodextrins advances the applicability of the dye for super-resolution imaging. Furthermore, the computational methods used here provide a rational way to search for better supramolecular hosts.

## MATERIALS AND METHODS

**Chemicals.** Cy3-Maleimide was purchased from GE Healthcare. β-cyclodextrin, methyl-β-cyclodextrin, and cucurbit[7]uril were purchased from Sigma-Aldrich Co. LLC. All the chemicals were obtained at the highest purity available and used without further purification.

**Computational modeling.** Crystal structures of α-CD, β-CD, γ-CD, and CB[n] (n = 5, 6, 7, and 8) were obtained from the Cambridge structural database.[52] Geometry of methyl-β-CD was generated through respective methyl substitutions to the β-CD, followed by geometry optimization using the Avogadro software (http://avogadro.openmolecules.net). Initial structure of Cy3 was generated in a similar manner, using the structure of a closely related molecule from cyanine family obtained from PUBCHEM database.[53] Cy3 was docked in CDs and CBs using the AutoDock Vina software,[37] with the search space limited to a 1.5 x 1.5 x 1.5 nm cube positioned at the centre of mass of the host. All rotatable bonds of Cy3 were allowed to be flexible during docking, while the host structure was confined to the initial crystal configuration. Ten best scoring conformations obtained from the docking were further energy minimized using the PM6-DH2 semi-empirical quantum mechanical energy model[38,39] and COSMO[54] continuum solvation model, in MOPAC software.[55] Interaction energies of the complexes were computed as the difference in energy of the complex and the sum of host and guest energies.

**Time-correlated single photon counting (TCSPC).** Diffusion coefficient, fluorescence lifetime, and molecular brightness were measured using a custom-built time-correlated single photon counting (TCSPC)[56] instrument equipped with an Olympus IX71 inverted microscope and a water-cooled 80 MHz, 5.4 ps, 75 mW, pulsed solid-state laser ($\lambda_{ex}$ = 532 nm, High-Q laser, Hohenems, Austria). A detailed description of our TCSPC instrument is documented elsewhere.[43] Briefly, the excitation light from the laser was passed through an appropriate filter and expanded to match the back-aperture of the objective. The laser was operated at low excitation power (< 100 µW) to avoid photobleaching and triplet state formation, and to primarily capture fluorescence intensity fluctuations arising from diffusion of molecules through the confocal observation volume. Emission light from the sample was passed through a dichroic mirror, a set of high-quality emission filters, and a polarizer fixed at the magic angle (54.7º), and collected using a 50µm/0.22NA optical fiber that served as a confocal pinhole. The emission photons were detected using a GaAsP photomultiplier tube, and the photon arrival times were recorded using the SPC-630 TCSPC module (Becker and Hickl, Germany).

Characteristic diffusion time of the fluorescent molecule was determined by auto-correlating the fluorescence intensity signal and fitting to a single-component three-dimensional diffusion model given by,[57,58]

$$G(\tau) = \frac{1}{N}\left[1+\left(\frac{\tau}{\tau_D}\right)\right]^{-1}\left[1+\left(\frac{1}{w}\right)^2\left(\frac{\tau}{\tau_D}\right)\right]^{-\frac{1}{2}}$$

where $N$ is the average number of diffusing fluorophores in the confocal observation volume, $\tau$ is the autocorrelation time, $w$ is the structure factor, and $\tau_D$ (= $r^2$/4D) is the characteristic diffusion time of the fluorophore with diffusion coefficient $D$ crossing a circular area with radius $r$. Non-linear least-squares fitting of the autocorrelation traces were performed using OriginLab 8.0 software. Radius of the observation volume determined from the known diffusion coefficient of Rhodamine 6G (D = 2.8 x $10^{-6}$ $cm^2$/sec in water) was ~ 400 nm and the structure factor was between 3 and 5 for all the experiments. Assuming that the diffusing fluorophore behaves as a hard sphere, hydrodynamic radius ($R_h$) was calculated from the diffusion coefficient using the Stokes-Einstein relationship,[59] $D = K_B T/6\pi\eta R_h$, where $K_B$ is the Boltzmann's constant, $T$ is the absolute temperature, and $\eta$ is the solvent viscosity (0.001 Pa.s for water). Molecular brightness was calculated as the ratio of the average fluorescence intensity signal to the average number of fluorophores ($N$) in the confocal observation volume.

Fluorescence lifetime was determined by fitting the histogram of photo arrival times (with respect to laser pulse time) to a bi-exponential decay model. Curve fitting was done using the FluoFit software (PicoQuant, Germany) by a process of iterative reconvolution given by the equation,[43]

$$I(t) = \int_{-\infty}^{t} IRF(t') \sum_{i=1}^{n} A_i e^{-\left(\frac{t-t'}{\tau_i}\right)} dt'$$

where $I(t)$ is the experimental decay function, $IRF(t)$ is the instrument response function, $A_i$ is the amplitude of $i^{th}$ lifetime, $\tau_i$ is the $i^{th}$ lifetime, and $n$ is the number of exponents (which is two in the case of bi-exponential decay model). The instrument response function (IRF) was collected from a sample of dilute scattering solution, prior to the experiment. Quality of the fitted curves was evaluated based on the autocorrelation of the residual curves. Average lifetime was calculated by intensity-weighting the individual lifetime components. All the measurements were taken at room temperature, with the dye and host concentrations at 10 nM and 1 mM, respectively, in deionised water.

**Steady-state spectroscopy.** Absorption and emission spectra of Cy3 in the free and complexed states in water were measured at a dye concentration of 3 μM, and a host concentration of 1 mM. Absorption spectra were obtained using the Agilent 8453 UV-Visible spectrophotometer (Agilent Technologies, Inc.), with a spectra bandwidth of 1 nm. Absorption spectra were corrected for the baseline value using the PeakFit software (Systat Software, Inc.) and absorption intensity at 532 nm was recorded. Emission spectra were obtained using RF-5301PC spectrofluorometer (Shimadzu, Inc.), with a spectra bandwidth of 1.5 nm. Emission spectra were collected at an excitation wavelength of 532 nm, and the emission wavelengths were restricted to the range of 540 nm to 700 nm. Quantum yields from the absorption and emission spectra were computed with Rhodamine 6G as reference, as described previously.[12,13]

**Photostability assay.** Photostability was characterized by monitoring the fluorescence intensity of the sample over time under constant epi-illumination.[13] The optical setup for epi-fluorescence microscopy was based on Olympus IX71 inverted microscope equipped with a 100 W Halogen lamp. Excitation light was passed through an appropriate filter before focusing in the sample through a 60x water-immersion objective (UPLAPO60XW, Olympus). Fluorescence emission was collected by the objective and passed through the emission filters to be detected by a high-sensitivity electron-multiplier CCD camera (SensicamEM, The Cooke Corporation, USA). Fluorescence images were captured after every 10 seconds for a total duration of 130 seconds, while the sample was continuously illuminated. Sample preparation involved sandwiching a small volume (~10 μl) of the dye solution between two cover slips. Fluorescence images were analyzed using ImageJ software (http://rsbweb.nih.gov/ij/). Non-overlapping regions-of-interest (ROI) of 10 x 10 pixels were randomly selected from three different regions of the sample for analysis. Photobleaching curves were generated by plotting the mean intensity of each ROI versus time.

## ACKNOWLEDGMENTS

Peter J. Butler gratefully acknowledges funding from the National Heart Lung and Blood Institute (R01 HL 07754201) and the National Science Foundation (BES 0238910). Ayusman Sen gratefully acknowledges funding by the National Science Foundation (DMR-0820404, CBET-1014673).

## REFERENCES

(1) Hell, S. W.; Wichmann, J.: Breaking the Diffraction Resolution Limit by Stimulated-Emission - Stimulated-Emission-Depletion Fluorescence Microscopy. *Opt Lett* **1994**, *19*, 780-782.
(2) Rust, M. J.; Bates, M.; Zhuang, X. W.: Sub-diffraction-limit imaging by stochastic optical reconstruction microscopy (STORM). *Nat Methods* **2006**, *3*, 793-795.
(3) Fernandez-Suarez, M.; Ting, A. Y.: Fluorescent probes for super-resolution imaging in living cells. *Nat Rev Mol Cell Bio* **2008**, *9*, 929-943.
(4) Yao, J.; Larson, D. R.; Vishwasrao, H. D.; Zipfel, W. R.; Webb, W. W.: Blinking and nonradiant dark fraction of water-soluble quantum dots in aqueous solution. *Proceedings of the National Academy of Sciences of the United States of America* **2005**, *102*, 14284-14289.
(5) Resch-Genger, U.; Grabolle, M.; Cavaliere-Jaricot, S.; Nitschke, R.; Nann, T.: Quantum dots versus organic dyes as fluorescent labels. *Nat Methods* **2008**, *5*, 763-775.
(6) Frangioni, J. V.: In vivo near-infrared fluorescence imaging. *Curr Opin Chem Biol* **2003**, *7*, 626-634.
(7) Ow, H.; Larson, D. R.; Srivastava, M.; Baird, B. A.; Webb, W. W.; Wiesner, U.: Bright and stable core-shell fluorescent silica nanoparticles. *Nano Lett* **2005**, *5*, 113-117.
(8) Larson, D. R.; Ow, H.; Vishwasrao, H. D.; Heikal, A. A.; Wiesner, U.; Webb, W. W.: Silica nanoparticle architecture determines radiative properties of encapsulated fluorophores. *Chem Mater* **2008**, *20*, 2677-2684.
(9) Santra, S.; Wang, K. M.; Tapec, R.; Tan, W. H.: Development of novel dye-doped silica nanoparticles for biomarker application. *J Biomed Opt* **2001**, *6*, 160-166.
(10) Santra, S.; Zhang, P.; Wang, K. M.; Tapec, R.; Tan, W. H.: Conjugation of biomolecules with luminophore-doped silica nanoparticles for photostable biomarkers. *Analytical chemistry* **2001**, *73*, 4988-4993.
(11) Saxena, V.; Sadoqi, M.; Shao, J.: Enhanced photo-stability, thermal-stability and aqueous-stability of indocyanine green in polymeric nanoparticulate systems. *J Photoch Photobio B* **2004**, *74*, 29-38.
(12) Morgan, T. T.; Muddana, H. S.; Altinoglu, E. I.; Rouse, S. M.; Tabakovic, A.; Tabouillot, T.; Russin, T. J.; Shanmugavelandy, S. S.; Butler, P. J.; Eklund, P. C.; Yun, J. K.; Kester, M.;

Adair, J. H.: Encapsulation of Organic Molecules in Calcium Phosphate Nanocomposite Particles for Intracellular Imaging and Drug Delivery. *Nano Lett* **2008**, *8*, 4108-4115.
(13) Muddana, H. S.; Morgan, T. T.; Adair, J. H.; Butler, P. J.: Photophysics of Cy3-Encapsulated Calcium Phosphate Nanoparticles. *Nano Lett* **2009**, *9*, 1559-1566.
(14) Fubini, B.; Hubbard, A.: Reactive oxygen species (ROS) and reactive nitrogen species (RNS) generation by silica in inflammation and fibrosis. *Free Radical Bio Med* **2003**, *34*, 1507-1516.
(15) Xia, T.; Li, N.; Nel, A. E.: Potential health impact of nanoparticles. *Annual review of public health* **2009**, *30*, 137-50.
(16) Lin, W.; Huang, Y. W.; Zhou, X. D.; Ma, Y.: In vitro toxicity of silica nanoparticles in human lung cancer cells. *Toxicol Appl Pharmacol* **2006**, *217*, 252-9.
(17) Brown, D. M.; Wilson, M. R.; MacNee, W.; Stone, V.; Donaldson, K.: Size-dependent proinflammatory effects of ultrafine polystyrene particles: a role for surface area and oxidative stress in the enhanced activity of ultrafines. *Toxicol Appl Pharmacol* **2001**, *175*, 191-9.
(18) Park, E. J.; Park, K.: Oxidative stress and pro-inflammatory responses induced by silica nanoparticles in vivo and in vitro. *Toxicol Lett* **2009**, *184*, 18-25.
(19) Dobrovolskaia, M. A.; McNeil, S. E.: Immunological properties of engineered nanomaterials. *Nature nanotechnology* **2007**, *2*, 469-78.
(20) Alexis, F.; Pridgen, E.; Molnar, L. K.; Farokhzad, O. C.: Factors affecting the clearance and biodistribution of polymeric nanoparticles. *Mol Pharmaceut* **2008**, *5*, 505-515.
(21) Barbe, C.; Bartlett, J.; Kong, L. G.; Finnie, K.; Lin, H. Q.; Larkin, M.; Calleja, S.; Bush, A.; Calleja, G.: Silica particles: A novel drug-delivery system. *Adv Mater* **2004**, *16*, 1959-1966.
(22) Bagwe, R. P.; Hilliard, L. R.; Tan, W. H.: Surface modification of silica nanoparticles to reduce aggregation and nonspecific binding. *Langmuir* **2006**, *22*, 4357-4362.
(23) Nau, W. M.; Mohanty, J.: Taming fluorescent dyes with cucurbituril. *Int J Photoenergy* **2005**, *7*, 133-141.
(24) Dsouza, R. N.; Pischel, U.; Nau, W. M.: Fluorescent dyes and their supramolecular host/guest complexes with macrocycles in aqueous solution. *Chemical reviews* **2011**, *111*, 7941-80.
(25) Koner, A. L.; Nau, W. M.: Cucurbituril encapsulation of fluorescent dyes. *Supramol Chem* **2007**, *19*, 55-66.
(26) Liu, S. M.; Ruspic, C.; Mukhopadhyay, P.; Chakrabarti, S.; Zavalij, P. Y.; Isaacs, L.: The cucurbit[n]uril family: Prime components for self-sorting systems. *Journal of the American Chemical Society* **2005**, *127*, 15959-15967.
(27) Moghaddam, S.; Yang, C.; Rekharsky, M.; Ko, Y. H.; Kim, K.; Inoue, Y.; Gilson, M. K.: New Ultrahigh Affinity Host-Guest Complexes of Cucurbit[7]uril with Bicyclo[2.2.2]octane and Adamantane Guests: Thermodynamic Analysis and Evaluation of M2 Affinity Calculations. *Journal of the American Chemical Society* **2011**, *133*, 3570-3581.
(28) Uzunova, V. D.; Cullinane, C.; Brix, K.; Nau, W. M.; Day, A. I.: Toxicity of cucurbit[7]uril and cucurbit[8]uril: an exploratory in vitro and in vivo study. *Org Biomol Chem* **2010**, *8*, 2037-2042.
(29) Rajewski, R. A.; Stella, V. J.: Pharmaceutical applications of cyclodextrins .2. In vivo drug delivery. *Journal of pharmaceutical sciences* **1996**, *85*, 1142-1169.
(30) Irie, T.; Uekama, K.: Pharmaceutical applications of cyclodextrins .3. Toxicological issues and safety evaluation. *Journal of pharmaceutical sciences* **1997**, *86*, 147-162.

(31) Arunkumar, E.; Forbes, C. C.; Smith, B. D.: Improving the properties of organic dyes by molecular encapsulation. *Eur J Org Chem* **2005**, 4051-4059.
(32) Ghale, G.; Ramalingam, V.; Urbach, A. R.; Nau, W. M.: Determining Protease Substrate Selectivity and Inhibition by Label-Free Suprannolecular Tandem Enzyme Assays. *Journal of the American Chemical Society* **2011**, *133*, 7528-7535.
(33) Hennig, A.; Bakirci, H.; Nau, W. M.: Label-free continuous enzyme assays with macrocycle-fluorescent dye complexes. *Nat Methods* **2007**, *4*, 629-632.
(34) Mishra, A.; Behera, R. K.; Behera, P. K.; Mishra, B. K.; Behera, G. B.: Cyanines during the 1990s: A review. *Chemical reviews* **2000**, *100*, 1973-2011.
(35) Thomson, J. M.; Parker, J.; Perou, C. M.; Hammond, S. M.: A custom microarray platform for analysis of microRNA gene expression. *Nat Methods* **2004**, *1*, 47-53.
(36) Polom, K.; Murawa, D.; Rho, Y. S.; Nowaczyk, P.; Hunerbein, M.; Murawa, P.: Current Trends and Emerging Future of Indocyanine Green Usage in Surgery and Oncology A Literature Review. *Cancer-Am Cancer Soc* **2011**, *117*, 4812-4822.
(37) Trott, O.; Olson, A. J.: AutoDock Vina: improving the speed and accuracy of docking with a new scoring function, efficient optimization, and multithreading. *Journal of computational chemistry* **2010**, *31*, 455-61.
(38) Korth, M.; Pitonak, M.; Rezac, J.; Hobza, P.: A Transferable H-Bonding Correction for Semiempirical Quantum-Chemical Methods. *J Chem Theory Comput* **2010**, *6*, 344-352.
(39) Rezac, J.; Fanfrlik, J.; Salahub, D.; Hobza, P.: Semiempirical Quantum Chemical PM6 Method Augmented by Dispersion and H-Bonding Correction Terms Reliably Describes Various Types of Noncovalent Complexes. *J Chem Theory Comput* **2009**, *5*, 1749-1760.
(40) Muddana, H. S.; Gilson, M. K.: Calculation of Host-Guest Binding Affinities Using a Quantum-Mechanical Energy Model. *J Chem Theory Comput* **2012**, *8*, 2023-2033.
(41) Muddana, H. S.; Gilson, M. K.: Prediction of SAMPL3 host-guest binding affinities: evaluating the accuracy of generalized force-fields. *Journal of computer-aided molecular design* **2012**, *26*, 517-25.
(42) Muddana, H. S.; Varnado, C. D.; Bielawski, C. W.; Urbach, A. R.; Isaacs, L.; Geballe, M. T.; Gilson, M. K.: Blind prediction of host-guest binding affinities: a new SAMPL3 challenge. *Journal of computer-aided molecular design* **2012**, *26*, 475-87.
(43) Gullapalli, R. R.; Tabouillot, T.; Mathura, R.; Dangaria, J. H.; Butler, P. J.: Integrated multimodal microscopy, time-resolved fluorescence, and optical-trap rheometry: toward single molecule mechanobiology. *J Biomed Opt* **2007**, *12*.
(44) Marquez, C.; Nau, W. M.: Polarizabilities inside molecular containers. *Angew Chem Int Edit* **2001**, *40*, 4387-+.
(45) Widengren, J.; Schwille, P.: Characterization of photoinduced isomerization and back-isomerization of the cyanine dye Cy5 by fluorescence correlation spectroscopy. *Journal of Physical Chemistry A* **2000**, *104*, 6416-6428.
(46) Sanborn, M. E.; Connolly, B. K.; Gurunathan, K.; Levitus, M.: Fluorescence properties and photophysics of the sulfoindocyanine Cy3 linked covalently to DNA. *Journal of Physical Chemistry B* **2007**, *111*, 11064-11074.
(47) Tabouillot, T.; Muddana, H. S.; Butler, P. J.: Endothelial Cell Membrane Sensitivity to Shear Stress is Lipid Domain Dependent. *Cell Mol Bioeng* **2011**, *4*, 169-181.
(48) Kanofsky, J. R.; Sima, P. D.: Structural and environmental requirements for quenching of singlet oxygen by cyanine dyes. *Photochem Photobiol* **2000**, *71*, 361-368.

(49) Buston, J. E. H.; Young, J. R.; Anderson, H. L.: Rotaxane-encapsulated cyanine dyes: enhanced fluorescence efficiency and photostability. *Chem Commun* **2000**, 905-906.
(50) Matsuzawa, Y.; Tamura, S.; Matsuzawa, N.; Ata, M.: Light-Stability of a Beta-Cyclodextrin (Cyclomaltoheptaose) Inclusion Complex of a Cyanine Dye. *J Chem Soc Faraday T* **1994**, *90*, 3517-3520.
(51) Montes-Navajas, P.; Garcia, H.: Cucurbituril Complexation Enhances Intersystem Crossing and Triplet Lifetime of 2,4,6-Triphenylpyrylium Ion. *J Phys Chem C* **2010**, *114*, 2034-2038.
(52) Allen, F. H.: The Cambridge Structural Database: a quarter of a million crystal structures and rising. *Acta crystallographica. Section B, Structural science* **2002**, *58*, 380-8.
(53) Wang, Y.; Xiao, J.; Suzek, T. O.; Zhang, J.; Wang, J.; Bryant, S. H.: PubChem: a public information system for analyzing bioactivities of small molecules. *Nucleic acids research* **2009**, *37*, W623-33.
(54) Klamt, A.; Schuurmann, G.: Cosmo - a New Approach to Dielectric Screening in Solvents with Explicit Expressions for the Screening Energy and Its Gradient. *J Chem Soc Perk T 2* **1993**, 799-805.
(55) Stewart, J. J. P.: Optimization of parameters for semiempirical methods V: Modification of NDDO approximations and application to 70 elements. *J Mol Model* **2007**, *13*, 1173-1213.
(56) Becker, W.: *Advanced time-correlated single photon counting techniques*; Springer: Berlin ; New York, 2005.
(57) Elson, E. L.; Magde, D.: Fluorescence Correlation Spectroscopy .1. Conceptual Basis and Theory. *Biopolymers* **1974**, *13*, 1-27.
(58) Magde, D.; Elson, E. L.; Webb, W. W.: Fluorescence Correlation Spectroscopy .2. Experimental Realization. *Biopolymers* **1974**, *13*, 29-61.
(59) Berg, H. C.: *Random walks in biology*; Expanded ed.; Princeton University Press: Princeton, N.J., 1993.